\documentclass[a4paper,12pt]{article}
\usepackage{hyperref}
\newcommand{\tr}{\mathop{\rm tr}\nolimits}

\def\a{\alpha}          \def\b{\beta}            \def\g{\gamma}
\def\d{\delta}          \def\eps{{\epsilon}}     \def\ve{\varepsilon}
                    \def\l{\lambda}
\def\m{\mu}             \def\n{\nu}              
                       
\def\vphi{\varphi}      \def\om{\omega}
\def\G{\Gamma}                     \def\L{\Lambda}
          
 \def\pa{\partial}
\def\be{\begin{equation}}            \def\ee{\end{equation}}
\def\ba#1{\begin{array}{#1}}         \def\ea{\end{array}}

\def\fr#1#2{\textstyle\frac{#1}{#2}}
  \def\cL{{\cal L}}

\begin{document}
\title{\bf Topological charges and quasi-charges \\
 in Absolute Parallelism}
\author{I. L. Zhogin\footnote{
 E-mail: zhogin at inp.nsk.su; \
 \href{http://zhogin.narod.ru}{http://zhogin.narod.ru }
} }
\date{}
 \maketitle

\begin{abstract}
The frame field theory, or Absolute Parallelism (AP), has
very many
interesting features: the large symmetry group of equations;
the field
irreducibility with respect to this group; variety of
differential covariants; and large list of consistent or
formally integrable second order equations not restricted to
Lagrangian ones.

There is one unique variant of AP (the unique equation,
non-Lagrangian) which solutions of general position seem to be
free of arising singularities if $D$=5. In the absence of
singularities, when degenerate (co)frame matrices 
are inaccessible and should be eliminated from the field set, AP
acquires the topological features of nonlinear sigma-model.

Starting with the
topological charge group, one can also introduce the concept of
 topological quasi-charge group for field configurations
having some symmetry. For $D$=4, considering symmetrical
equipped 0-(sub)manifolds
 in ${\bf R}^3$,
 we calculate quasi-charge groups $\Pi(G)$ for a number
of symmetry groups $G\subset O_3$, and describe morphisms of
$\Pi$-groups.

Then, the differential 3-form of topological charge, which is
dual to the topological current $J^{\mu}$, is derived, as well as
 the 1-form of topological quasi-charge for SO${}_3$--symmetric
field configurations.

The problem for $D$=5 is briefly discussed, and results of
topological classification of symmetric field configurations
(alighting on evident parallels with the Standard Model
combinatorics of fundamental particles)
are announced. An example of $SO_2-$symmetric
 configuration is considered and the  quasi-charge
3-forms -- both `left' and `right' (or self-dual and
anti-self-dual) -- are obtained (as well as the 4-forms of
topological charge).

In conclusion, we propose a variant of experiment
with single photon interference (or with bi-photon
non-local correlations) which should verify
a possible non-local (spaghetti-like) 5$D$ ontology
of particles.

\end{abstract}


\section{Introduction}
Most (if not all) modern works on Absolute Parallelism
(either pure AP, or modified with extra structures) [1--5]
 follow the Lagrangian approach to obtain field equations
 (Ref.\ \cite{sa} also has many historical comments on AP).
However, the large list of consistent second order equations of
 AP discovered by Einstein and Mayer \cite{eima}
 (the earlier Einstein's original papers on AP
 are  available in English translation \cite{uc})
 includes also three classes of non-Lagrangian equations;
one of these classes is of particular interest and it admits
3-linear  presentation
 \cite{zh3,zh4}.

After due consideration for AP equations and notations used in the
paper, we will return to introduction notes. Next, the main goal
of this paper will be to demonstrate that spatially localized
field configurations of AP can carry integer data --
topological charge and/or quasi-charge (in the case of
symmetrical field configurations), and to show how to calculate
these charges. (Our tacit intention is also to outline
 conceivable phenomenology of such particle-like-configurations,
indicating parallels with the existing combinatorics
of elementary particles and asking ourself whether this
phenomenology can benefit (in some circumstances)
from all the machinery of quantum
(field) theory -- including principle of superposition, and
path integrals, and so on.)
In the last section a variant of experiment
with single photon interference (or bi-photon
non-local correlations) is suggested which might verify
a possible non-local (spaghetti-like) 5$D$ nature
of `particles'.
\subsection{Notations and consistent AP equations}
 In this paper we will be dealing with the only field --
 frame field $h_a{}^\mu$, and inverse matrix --
 co-frame field
 $ h^a{}_\mu(x^\nu)$; this field defines the
  space-time metric as usual
 using the Minkowski metric:
\begin{equation}\label{gmunu}
 g_{\mu \nu }= \eta_{ab}h^a{}_\mu h^b{}_\nu \, ,
 \mbox{ where }
 \eta_{ab}=\eta^{ab}=\mbox{diag}(-1,1,\ldots,1).
\end{equation}

We will use coma "," and semicolon ";" to denote partial
derivative and usual covariant differentiation with  symmetric
Levi-Civita (or Christoffel) connection, respectively, as well
as the following tensors (differential covariants):
\begin{equation}\label{gal}
  \gamma_{a\mu\nu} = h_{a\mu;\nu}\ , \
\ \ \Lambda_{a\mu\nu} = 2\gamma_{a[\mu\nu]}=h_{a\mu,\nu}-
h_{a\nu,\mu}\ .\end{equation}
 We shall omit in contractions the
matrices $\eta^{ab}, \eta_{ab}$ and,
 in covariant expressions (where only the covariant
 differentiation is in use),
-- $g^{\mu\nu}, g_{\mu\nu}$, because
\[
 0=g^{\mu\nu}{}_{;\lambda}=g_{\mu\nu}{}_{;\lambda}
 = \eta^{ab}{}_{,\mu}= \eta_{ab}{}_{,\mu}.
\]
Let  introduce notations  for the following (irreducible)
covariants:
  \begin{equation} \label{s-f}
  S_{\mu\nu\lambda} = 3 \Lambda_{[\mu\nu\lambda]}, \
  \Phi_\mu = \Lambda_{a a\mu}, \
  f_{\mu\nu} = \Phi_{\mu;\nu} - \Phi_{\nu;\mu}\, .
 \end{equation}
The type of an index is changed by means of (co)frame, so the
same letter is used for covariants with any indices -- Latin, or
Greek, or mixed, with the only evident exception for (co)frame,
metric, and Minkowski matrix.
 For example:
 \[
\g_{\mu\nu\lambda}= h_{a\mu} \g_{a\nu\lambda} \
 (= \g_{[\mu\nu]\lambda}  ), \
 f_{ab}=f_{\mu\nu}h_{a\mu}h_{b\nu}.
\]

Note that the definition (\ref{gal}) leads to the following
identities:
 \be \label{ide}
 \Lambda_{a[\mu\nu;\lambda]} \equiv 0\,,
  \ \  h_{a\l}\Lambda_{abc;\l}+f_{bc}\equiv 0\, .
  \ee

The most simple case of consistent (or formally integrable, or
well-posed) AP equations is the two-parameter non-Lagrangian class
II$_{22112}$ of
  \cite{eima} (the Lagrangian equations should have the term
  $h_{a\mu} \cL$):
\begin{equation}
\label{sys1} {\bf A}_{a\mu}=K_{a\mu\nu;\nu}=0, \
K_{abc}=K_{a[bc]}=\a\Lambda_{abc}-\b S_{abc}-\g
\eta_{ab}\Phi_{c}   +\g \eta_{ac}\Phi_{b}\,;
\end{equation}
here the overall coefficient is an arbitrary value.
The evident identity
\[ {\bf
A}_{a\mu;\mu}\equiv 0 ~~(K_{a\mu\nu;[\nu;\mu]}\equiv 0) \]
ensures the consistency of the Eq.~(\ref{sys1})
with the exception of ``bad''\
cases: $\a{=}0$, or
 $2\b{=}\a$, or (and) $\g{=}\a$;
 it is preferable to place $\a{=}1$ and
this will be in effect below.

(In the case $\a{=}0$, the prolonged equation
${\bf A}^*_{\mu(\nu\l)}= {\bf A}_{(\mu\nu);\l}
+{\bf A}_{(\mu\l);\nu}-{\bf A}_{(\nu\l);\mu}=0$
gives: $\Phi_{\mu;(\nu;\l)}- O(\L\L',\L^3)=0$,
and the next prolonged
equation,
${\bf A}^*_{\mu(\nu\l);\tau}-{\bf A}^*_{\mu(\nu\tau);\l}
= 0 $,
leads to a new and irregular equation:
the principal derivatives vanish, but `quadratic' terms do not.
The other `bad' cases will be explained below.)

Einstein and Mayer had usually used not Levi-Civita, but
Weitzenb\"ock connection derived from the condition
\begin{equation}\label{weitzen}
 0=h_{a\m\hat{;}\n}=
 h_{a\m,\n}-\hat{\G}^\l_{\m\n}h_{a\l};
 \ \ \hat{\G}^\l_{\mu\nu}=h_{a}{}^{\l}h^a{}_{\m,\n}.
\end{equation}
 However, the symmetric connection is equally suitable
 to write any covariant system of AP (in this case,
 `Maxwell-like' equations for
 skew-symmetric covariants like in the Eq.~(\ref{sys1})
  become even more clear).

The next evident class (labelled in \cite{eima} as I$_{12}$)  of
consistent equations is the two-parameter class of Lagrangian
equations. Using the scalar density
 $h{\cal L}, \ h=\det
h^{a}{}_{\mu}$, with two free parameters ($\a{=}1$),
\[ {\cal L}=\fr 1 4 \Lambda_{abc}\Lambda_{abc} -
\fr \b {12} S_{abc}S_{abc} - \fr \g 2 \Phi_{a}\Phi_{a}~,\]
 and taking into account the
 symmetry properties of $\Lambda$ and $S$,
 see (\ref{gal}), (\ref{s-f}), and (\ref{sys1}),
 one can obtain the Lagrangian equations:
\[d{\cal L}= \fr1 2 K_{abc}\, d\Lambda_{abc}=
K_{a}{}^{\m\n}\, dh^a{}_{\mu,\nu}
- \Lambda_{bca}K_{bc}{}^\mu \, dh^a{}_\mu  ~, \]
\begin{equation}
\label{sys2} {\bf B}_a{}^\mu= - \frac{\delta
(h{\cal L})}{h\,\delta h^a{}_\mu} =K_a{}^{\mu\nu}{}_{;\nu}
+\Lambda_{bca}K_{bc}{}^\mu -h_a{}^\mu{\cal L}=0~.
\end{equation}
The identity providing the formal integrability looks as follows:
\[ {\bf B}_{a\mu;\mu } -{\bf B}_{bc}\Lambda_{bca} \equiv 0\, ; \]
 so, we have the conservation law:
 $(h\Lambda_{bca}K_{bc}{}^{\mu }
 -hh_{a}{}^{\mu}{\cal L})_{,\mu}=0 $.
 The skew-symmetric part of (\ref{sys2}),
 \[
2{\bf B}_{[\mu\nu ]}=(1 -2\b)(S_{\mu \nu \lambda ;\lambda }
-\Lambda_{\mu ab}\Lambda_{ab\nu } +\Lambda_{\nu ab}\Lambda_{ab\mu
})+ (1-\g) (f_{\mu \nu }-\Phi _a\Lambda_{a\mu \nu })=0, \]
 disappears (or turns into identity) if
  $2\b{=}\g{=}1$, and this is the case of the General Relativity
 (of course, in this case one (do) can add
 a skew-symmetric equation with a large arbitrariness).

 The cases $2\b{=}1,\,\g{\neq}1$ and $\g{=}1,\,2\b{\neq}1$
 are  both ``bad" (ill-posed), because the equation
 ${\bf B}_{[\mu\nu ;\lambda ]}=0$, or, respectively,
 ${\bf B}_{[\mu\nu ];\nu}=0$, leads to a new and irregular
 second order equation. The last case,
 despite its incorrectness, still can be found in
 publications\cite{auon,nashed}.

 The third class {\bf C} (two-parameter class labelled
 in  \cite{eima} as  I$_{221}$) is
 possible (that is, consistent) only in 4$D$, and its equations
 in symmetric part resemble the equations
  of  Brans-Dicke theory (with $\Phi_\m$ instead of
  $\phi_{,\m}/\phi$; here $G_{\m\n}$ is the Einstein
  tensor)
 \[
 {\bf C}_{(\mu \nu )}=
 2G_{\mu\nu}+2\tilde{\g}(\Phi_{(\mu;\nu)}
-g_{\mu\nu}\Phi_{\tau;\tau})
+\tilde{\g}^2(\Phi_\mu\Phi_\nu+
g_{\mu\nu}\Phi^2/2)=0,\]
 while the skew-symmetric part is
  quite simple:
 ${\bf C}_{[\mu \nu ]}=S_{\mu \nu \lambda ;\lambda }
+ \tilde{\b}f_{\mu \nu }=0$ \cite{the}.

The last and most interesting class of  consistent AP equations
(the one-parameter class II$_{221221}$ of \cite{eima}) admits the
next presentation
(here $ L_{a\mu \nu }=L_{a[\mu \nu]}=
\Lambda_{a\mu \nu }-S_{a\mu \nu }-2\g h_{a[\mu }\Phi_{\nu]}$):
 \begin{equation}
\label{sys4} {\bf D}_{a\mu}=L_{a\mu\nu;\nu}- (1-2\g ) (f_{a\mu
}+L_{a\mu \nu }\Phi _{\nu })=0\, ,
\end{equation}
\[
\!\!\!\!\!
\mbox{and \ }
2{\bf D}_{[\nu\mu]}=S_{\mu\nu\l;\l}+ (1-3\g )
(f_{\mu\nu}-S_{\mu \nu\l }\Phi _{\l })\, . \]
The equations ${\bf D}_{a\mu;\mu}=0$ and
${\bf D}_{[\mu\nu];\nu}=0$ give the same
 `Maxwell equation' (hence, the identity required for consistency
 is here):
\begin{equation}\label{max}
 f_{\mu\nu;\nu}=(S_{\mu \nu\l }\Phi _{\l })_{;\l} \ \
(= (1-3\g )f_{\mu\nu}\Phi _{\n }
-\fr1 2 S_{\mu \nu\l }f_{\n\l})
\, .
\end{equation}

 The case $\g{=}\fr13$ of Eq.~(\ref{sys4})
 is of special interest, and it
{\em requires\/} extra-dimension(s) because the trace equation
 ${\bf D}_{aa}=0$ becomes irregular (the principal
 derivatives vanish) if $D=1+\g^{-1}$.


\subsection{Introduction, continued}

 We believe  that the frame field theory
 (aka Absolute Parallelism)
  is to become of
significant interest for mathematical physics (as well as for
cosmology and particle physics; better to say -- for physics)
because of the following reasons and features:

\begin{enumerate}
    \item{High symmetry of this theory:}

This symmetry group includes symmetries of both special and
general relativity theories; so, staying in
a class of equivalent
 solutions, $h(x)$, to some AP equation, one
 can perform diffeomorphic coordinate mappings
  (democracy of ``good'' coordinates) and global
 transformations from ``completed'' Lorentz group
 (this is the point symmetry group of inertial coordinates,
 which includes also the global scale transformations):
 \begin{equation}\label{hsym}
 \tilde{h}^{a}{}_\mu(\tilde{x})
 =\kappa\, \sigma^a{}_b h^b{}_\nu(x) \frac{\partial
 x^\nu}{\partial \tilde{x}^\mu}~,
 \label{tran}
\end{equation}
where $\kappa>0$, $\sigma^a{}_b \in O(1,D{-}1)$;
$\kappa, \sigma^a{}_b =\;$const.
 The irreducibility of
 this (vector) field representation is also very important
feature helping to avoid free parameters.\footnote{In
 gage theories, e.g., the need to combine a gage field and matter
ones (reducibility) leads immediately to an arbitrary parameter --
the gage constant.}

In our opinion, the space-time signature without Lorentz-group (as
it takes place in GR) looks  a bit like beer-foam without beer,
 or (for very versed specialists) like the smile without the cat;
 it means that the choice of signature in GR is the separate
postulate having no relation to the symmetry group of GR.

On the other hand, the inertial coordinates of Special Relativity
(which are the basis of all QFT), being although simple and
aristocratic, have some of strangeness and unnaturalness. Both
AP  and  GR  say
 (answering the Mach's question) that
 the space-time geometry is not an immutable essence, and that
 the inertial coordinates,
  $y^a$, are
 the property (or attribute) of trivial solution only.
  Indeed, one can integrate the equations
  $ y^a{}_{,\mu} = h^a{}_\mu $,   if  $\Lambda_{a\mu\nu}=0.$
  Are these coordinates not a bit an empty, unreachable abstraction
  scarcely suitable to be a solid basis for a really fundamental
  theory?

  The \emph{Diff-}covariance is of practical importance as it gives
  another sense to the gradient catastrophe problem
  (solution becomes being multivalued) -- as having relation to
  co-singularities
  (naturally, this problem is not burning if you are in the perturbative
domain); at the same time the field irreducibility reduces
or restricts the number
of ways to singularities of solutions:
  the rank of frame (co-frame) matrix is the only significant
  parameter for dealing with (co-) singularities (compare with bi-
  or non-symmetric metric).
   \item

  Absence of arising singularities and uniqueness:

There is one unique variant of AP (non-Lagrangian, with the unique
$D$; $D$=5) which solutions of general position seem to be free
of arising singularities.
 Extension of formal integrability test (applied
for second order AP-equations) to the cases of degeneration of
either co-variant frame matrix (co-singularities) or
contra-variant frame (or contra-frame density of some weight)
may serve, we believe, as the local and covariant (no coordinate
choice) test for singularities of solutions. In AP this test
singles out the unique equation,
 the case $\g{=}\fr13$ of the Eq.~(\ref{sys4}),
  and the space-time dimension
mentioned above \cite{the,zh5,zh3}.

Solutions to this equation, if they exist not only 20 billion
years, and not only septillion years, but unlimited time, always
(there are no stops with singularities), still being `interesting',
should be very complicated having many parameters and scales
(the unique equation itself has no characteristic scales).
Very slowly varying parameters might serve for some fast
processes as `phenomenological constants', and it might be very
useful to develop various, sighted phenomenologies or simplifying
models.

``Any change of the fundamental theory should destroy this
theory" -- this Einstein's expert estimation (somewhere in his
autobiographical notes) could serve as a principle or ideal,
the principle of uniqueness.  So, this ideal of oneness seems to be
achievable in AP, if Nature does not like singularities (preferring
that the single standard of regularity property to be valid for all
space-time points).

    \item 
   Energy-momentum tensor and weightless (or energy-less)
      waves: 

Although non-lagrangian, the unique equation leads to the
symmetric, ``covariantly conserving" energy-momentum tensor
where the main (second order) terms  depend only on
the rank two tensor
 $f_{\mu\nu}$, which looks like
electromagnetic field (however, there are no gradient
transformations in the theory).
Applying the field equations (\ref{sys4}), and prolonged
equations as well, to
$h^{-1}\, \delta(h\,R_{\mu\nu}G^{\mu\nu})/\delta g_{\mu\nu}$
($R,G$ are the Ricci and Einstein tensors), one can find
$T^{\mu\nu}$ \cite{the,zh5}.
 In the theory, there are
 solutions with $f$=0, see Eq.~(\ref{max}),
  which carry no (or almost no)
 energy-momentum, nor angular momentum.

 So, in $D$=5, in weak field approximation
 only three of 15 polarizations (i.e. plane wave solutions) do
 carry energy-momentum, while the others do not.
  This is a very unusual feature -- impossible
 in the Lagrangian tradition; it needs some time,
 week or month, that to become used to this `singularity'.

 One can show \cite{the,zh5} that a wave-packet
of the energy-carrying $f$-component  should move along usual
Riemannian geodesics -- as in GR (if background has $f$=0);
however, the polarization (or spin)  evolution should depend
also on the rank three skew-symmetric tensor $S_{\mu\nu\lambda}$,
which is certainly absent in GR.

Another strange feature is the instability of trivial solution:
 some weightless polarizations grow linearly with time in presence of
 ponderable $f$-polarizations. Really, from the linearized
 Eq.~(\ref{sys4}) [and the identity (\ref{ide})] one can write
 (the following equations should be treated as linearized):
 \[ \Phi_{a,a}=0 \ (D\neq1+\gamma^{-1}), \
 \Lambda_{abd,d}=\gamma(\Phi_{a,b}-2\Phi_{b,a});\]
 \[\Lambda_{a[bc,d],d}\equiv0; \
 \Lambda_{abc,dd}=-2\gamma f_{bc,a}\, .\]
 The last `D`Alembert equation' has the `source' in its right
 hand side. Some components of $\Lambda$
 (most symmetrical irreducible parts)
 do not grow (as well as Riemannian curvature), because
 (again, linearized equations are implied below)
 \[ S_{abc,dd}=0, \ \Phi_{a,dd}=0 \ \,
  (\mbox{also } f_{ab,dd}=0, \ R_{abcd,ee}=0), \]
  but the least symmetrical components of the tensor $\Lambda$
  do go up (with time $t$ -- due to terms
   \ $\sim t\, e^{-i \om t}$)
  if ponderable waves, the three $f$-polarizations
  are not vanishing.

 \item 
  Non-stationary spherically symmetric solutions as expanding
cosmological model:

 The unique symmetry of AP equations gives scope for symmetrical
 solutions.
Non-stationary spherically ($O_4$-)
 symmetric solutions to the 5$D$ unique equation
 lead through a number of integrations to
 specific scalar fields which can serve as privileged
 radius and time (quasi-inertial coordinates;
 this looks more suitable to match with observable cosmology).
  The condition $f_{\mu\nu}{=}0$ is a must for solutions with such
 a high symmetry (as well as
 $S_{\mu\nu\l}{=}0$); so, these $O_4$-symmetric solutions
 carry no energy, that is, weight nothing
 (some lack of \emph{gravity} !\footnote{%
It seems that the universe
 expansion has little common with  GR
 and its dark energy.}).

 More realistic cosmological model may look like a single
 $O_4$-symmetric wave
 (or a sequence of such waves) moving along the radius and being
 filled with a noise, or  stochastic waves both
 weighty (\emph{weak}, $\Delta h\ll1$) and
  weightless ($\Delta h<1$, but intense enough that
  to lead to non-linear
  fluctuations with $\Delta h\sim1$)
  which form  statistical ensemble(s)
 having a few characteristic parameters
 (after `thermalization').
  The development and examination of stability
 of such a model is an interesting problem. One may
 think that $O_4$-wave (of proper sign) can serve as
 a time-dependent `shallow dielectric guide' for that weak
noise waves. The ponderable  waves (noise-1) should
have wave-vectors almost
tangent to the $S^3$-sphere of wave-front that to be trapped
inside
this (`shallow') wave-guide; the imponderable waves (noise-2)
can grow up, and escape from the wave-guide (due to
non-linear effects of scattering), and their wave-vectors
can be less tangent to the sphere.

The waveguide thickness can be small for an observer in
the center of $O_4$-symmetry (may be, some larger than
inverse `temperature', or the end of noise spectrum,
$\l_n$), but in co-moving system
it can be very large (due to relativistic effect), however
still small with respect to the radius of sphere,
$L\ll R$. It seems that the radial dimension has to be
 very `undeveloped'; that is, there are no other characteristic
scales, smaller than $L$, along this extra-dimension.

    \item 
 Topological features of  non-linear sigma-model:

 In the absence of solutions (of general position) with arising
singularities, when degenerate (co- and contra-)
frame matrices
are unreachable and can be eliminated from the field set, AP
acquires topological features of non-linear
sigma-model.\footnote{This way to topological
(quasi)charges looks more natural than
all the `crystal spheres of chiral models'.}

Starting with topological charge group, one can introduce
further the concept
 of topological quasi-charge group for field
 configurations having some symmetry \cite{the}.
It seems that the possible variety of quasi-charges in 5$D$ case
 (living on
the ``cosmological background" of O$_4$-symmetric solution
filled with weak stochastic waves) could be sufficient to
explain qualitatively many (if not most of) features of the Standard
Model (including superposition principle and Feynman's path
integral -- as a result of integration over 5-th dimension in
``cosmological waveguide").

The non-linear, particle-like field
configurations with quasi-charges (quasi-particles) should be
very elongated along the extra-dimension
(all of the same size $L$),
while being small sized along usual dimensions, $\l_n\ll L$.
The motion of such a spaghetti-like quasi-particle should be
very complicated and stochastic due to `strong' imponderable
noise, such that different parts of spaghetti are coming
their own paths.
At the same time, quasi-particle can acquire
`its own' energy--momentum only due to scattering of ponderable
waves (which  wave-vectors are almost tangent to usual 3$D$
(sub)space);
so, it seems that scattering
amplitudes\footnote{%
These amplitudes can depend on additional vector-parameters
(`equipment vectors') relating to differential of field mapping  at a
`quasi-particle center' -- where quasi-charge density is
largest (if it has covariant sense).}
of those spaghetti's
parts which have the same 3$D$--coordinates can be summarized
providing an auxiliary, secondary
[or shadow,\footnote{Note some similarity
(except superposition, of course)
with the Plato's `shadows in a cave'.}
(1+3)$D$] field; non-linear
fluctuations of noise-2 are responsible for `zero-point
oscillations' of secondary fields.

So, the imponderable noise-2 provides stochasticity
(of motion of spaghetti's parts), while
the ponderable noise-1 gives superposition (with secondary fields).
Phenomenology of secondary fields could be of Lagrangian type,
with positive energy acquired by quasi-particles, -- that to
ensure the stability (of all the waveguide with its infill
with respect to quasi-particle production; the least action
principle has deep concerns with Lyapunov stability and is
deduced, in principle, from the path integral approach).
\end{enumerate}

Thus, we believe that there has to be a definitive difference
between the
right theory and all the others, wrong ones.
We're expecting the right
theory to be of `monotheistic' kind --
 with no free parameters and no room
for changes. (Some people believe that such a simple
theory is impossible\cite{schroer} and our world is
infinitely complex -- but according to Einstein, this case `is not
interesting'.)

A theory with arbitrary parameter(s) looks like `fire on areas'.
We really need only one fundamental theory, but not a heap of
slightly different theories where only one especial variant is
supposed to be the Truth (while we have no good way to distinguish
this Truth from Untruth -- during limited experimental
time and using limited
funding; on the other hand, theories with free parameters are
quite appropriate for the maxim 'the show must go on').


\section{Topological charges and quasi-charges}
 In this section the topological properties of spatially
 localized configurations of frame field,
$h^{a}{}_{\mu}\to \delta^{a}_{\mu}$
 on space-like infinity,  are to be explored.

 We will suppose that Riemann space defined by metric is of
 trivial topology: no worm-holes, no compactified space
 dimensions, no singularities. A process of
 (de)compactification would require arising singularities which
 are thought to be impossible
 (due to careful choice of field equations),
 and we prefer that the single standard
 of no-compactification to be applied
 to all space(time) dimensions.

 It is possible to continuously
deform the metric and, simultaneously, the frame field
 $h(x)$  such that the metric
  becomes trivial,
equal to Minkowski metric,
$g_{\mu\nu}\to\eta _{\mu\nu} $,
whereas $h$-field becomes a 
field of rotation matrices (metric can be
subjected to diagonalization,\cite{schutz}
 `square-rooting', and so on;
we introduce here the notation: $m{=}D{-}1$)
\begin{equation}\label{deform}
h^{a}{}_{\mu}(x)  \to s^{a}{}_{\mu}(x)\in SO(1,m).
\end{equation}
Further deformation is possible that to remove boosts too,
and so, for
any space-like (Cauchy) surface, this gives a (pointed) map:
\[ s\,: \ {\bf R}^m\cup \infty=S^m \to SO_m;
\ \infty\mapsto 1^m\in SO_m .\]
The set of such maps consists of homotopy classes which
form an Abelian group,
 and this is the group
of topological charge, $\Pi(m)$:
\begin{equation}\label{pim}
\Pi(m)=\pi_m(SO_m); \ \ \Pi(3)=Z, \ \Pi(4)=Z_2+Z_2.
\end{equation}
Here $Z$ is the infinite cyclic group, and
$Z_2$ is the cyclic group of order two (two group)
\cite{du}.

Due to great symmetry of AP-equations, see (\ref{hsym}),
their solutions also can possess some symmetries, which should
be a diagonal subgroup of that great symmetry.
It is very important that the deformation to $s$-field
can be performed such that to keep  symmetry of field
configuration.
\subsection{Quasi-charge groups}
 We are going to give (a sketch of)
 topological classification of symmetric
 localized configurations of $SO_m$-field.
Let us repeat the assumed definition: localized field
(or pointed map)
 $$  s(x) :\  {\bf R}^{m} \to SO(m),\ \  s (\infty)=1^{m},$$
is $G$-symmetric if, in some suitable coordinates,
\begin{equation}
\label{gsi}
  s(\sigma x)= \sigma s (x)\sigma^{-1}\
 \ \forall \ \sigma \in G \subset O(m)\, .
\end{equation}
 The set of such fields ${\cal C}^{(m)}_G $ generally consists of
 separate, disconnected components -- homotopy classes forming
the Abelian group which is denoted here as  $\Pi (G;m)$; i.e.
 $$ \Pi (G;m)\equiv\pi_{0}({\cal C}^{(m)}_G) \, .$$
 For such a group we will coin the term `topological quasi-charge
 group', or QC-group.

These groups also classify symmetrical localized
configurations of frame field $h^a{}_\mu(x)$ as it was
outlined above.
 Since field equation does not break symmetry,
 field configuration with non-trivial quasi-charge merits
 some good name
 (something better than), say, \emph{topological quasi-soliton},
 or {\em quasi-particle}.
If symmetry is not exact (because of some distant regions),
quasi-charge is not exactly conserving value, and quasi-soliton
(of zero topological charge) can vanish or arise
 during colliding with another quasi-soliton.

 Along with calculation QC-groups for different symmetries
  we should solve
 another problem. Let $G1$ includes $G2$ (with respect
 to its elements or generators), such that there is a
 mapping (embedding) of field configurations:
 $$  i:\ {\cal C}^{(m)}_{G1} \to {\cal C}^{(m)}_{G2}.$$
This mapping induces the homomorphism of QC-groups:
$$     i_{*} :  \   \Pi(G1;m) \to  \Pi(G2;m), $$
so one has to describe this and to find how the `small' pieces
 of more symmetric fields  are situated within
 the  `large lumps' of less symmetric
 field configurations. 
\subsection{Examples of QC-groups: diad (and $k$-ad)
homotopy groups }
Let us consider the very simple (discreet) symmetry group  $P_1$
with a plane of reflection symmetry:
 $$P_1 =\{1,p_{(1)}\},
\mbox{ \ where } p_{(1)}=
\mbox{diag}(-1,1,\ldots,1)=p_{(1)}^{-1}.$$

 It is necessary to set field $s (x)$ on the half-space $\frac1
2\,{\bf R}^m=\{x^1\geq 0\}$, with additional condition imposed on the
surface ${\bf R}^{m-1}=\{x_1=0\}$
(stationary points of $P_1$ group) where $s $ has
to commute with the symmetry [see (\ref{gsi})]:
\[p_{(1)}x=x\ \Rightarrow\ s(x)=p_{(1)}s p_{(1)}
\ \Rightarrow\ s\in 1 \times SO_{m-1}. \]
 Hence, accounting for the localization requirement,
 we have a diad map
 (relative spheroid; here $D^m$ is an $m$-ball and
 $S^{m-1}$ its surface)
\[(D^m;S^{m-1}) \to  (SO_{m}; SO_{m-1}),\]
and topological classification of such maps leads to
the relative (or diad) homotopy group\cite{du}:
\[\Pi(P_1;m)=\pi_{m}(SO_{m};SO_{m-1})
=\pi_{m}(S^{m-1}). \]
 The last equality follows due to fibration
$SO_{m}/SO_{m-1}=S^{m-1}$, also see \cite{du}.

Similar considerations (of group orbits
and stationary points) lead to the following result:
\[\Pi(O_{l};m)=\pi_{m-l+1}(SO_{m-l+1};SO_{m-l})
=\pi_{m-l+1}(S^{m-l}). \]
If $l>3$, there is the equality: $\Pi(SO_{l};m)=\Pi(O_{l};m)$,
while for $l=2,3$ one can find\cite{the}:
\[
\Pi(SO_{3};m) = \pi_{m-2}(SO_{2}\times SO_{m-2};SO_{m-3})=
\pi_{m-2}(S^1\times S^{m-3}),
\]
\[
\Pi(SO_{2};m) = \pi_{m-1}(SO_{m};SO_{m-2}\times SO_{2})=
\pi_{m-1}(RG_+(m,2)).
\]

At last, the `compound' symmetries,
$G=G_1\times\cdots\times G_k$ (all $G_i$ are simple),
having more complicated picture of (sub)orbits, and
stationary points (for every $G_i$), and their intersections,
lead to $(k+1)-$ad homotopy groups, which can be defined
(by induction\cite{the}) as a generalization of diad\cite{du}
and triad\cite{blak} homotopy groups.
\section{QC-groups, their morphisms and forms in 4$D$}
In this section $m=3$, and it will be omitted from $\Pi$-groups.
The Pontryagin's method of substitution of maps with equipped
sub-manifolds \cite{po,du} can be extended for the case of
$G-$symmetric maps (\ref{gsi}). This gives the useful way to
find both quasi-charge groups and their morphisms.
For example, one can show that an
equipped  point
(say, positive; that is, a three vectors equipment has
 the positive orientation)
can be placed at the center of $SO_3-$symmetry, or at the line
of (stationary points of) $SO_2-$symmetry. It means that there
are isomorphisms (with the group of topological charge):
\[ \Pi(SO_3)\cong\Pi(SO_2)\cong\Pi=Z .\]

Let's briefly illustrate this approach using the example of
discreet symmetry groups $P_2\{1,2\}$
(inversion of first two
space coordinates, or their $\pi-$rotation)
 and $P=P_3$ (full space inversion).
One can find that
\[\Pi(P_2\{1,2\})=\Pi(P_2)=
Z_{\rm(a)}+Z_{\rm(p)},\]
where the first $Z$ describes the equipped points
(the difference of positive and negative points)
at the axis
of symmetry (the third axis, where $x^1=x^2=0$),
while the other $Z$ describes the pairs of
peripheral points (number of positive pairs minus
negative; $\pi-$rotation does not change the sign
of equipment orientation). So, one can describe
the morphism as well (isomorphism plus monomorphism-2):
 \[ \Pi(P_2)\supset Z_{\rm(a)}\ni 1
 \mapsto 1\in Z=\Pi\, , \ \
Z_{\rm(p)}\ni 1 \mapsto 2\in Z=\Pi\,.
 \]
In the case of $P_3$, any equipped point should have
a partner of opposite sign in the inverted position
(the inversion change the sign of equipment, and the
topological charge is a pseudoscalar), but one can
bring any two such pairs into annihilation keeping the
symmetry. All that means that
 \[ \Pi(P_3)=Z_2\ni 1 \mapsto 0\in Z=\Pi\, . \]
Similar consideration gives (also zero morphism)
 \[\Pi(P_1)=Z\ni 1 \mapsto 0\in Z=\Pi\, . \]
\subsection{Differential 3-form of topological charge (in 4$D$)}
The topological current for a spatially localized field
of rotation matrices
$s(x^{\mu}):\ {\bf R}^{1+3}\to SO_3$ (or $\to SO(1,3)$)
 may be defined as follows
 (much as the topological current in the
  Skyrme model with $SU_2$-field \cite{skyrme,rybakov,ward})
\[
J^\mu _{(s)}\propto\varepsilon^{\mu\nu\varepsilon\tau }
 \tr(
\gamma^*_\nu\gamma^*_\varepsilon\gamma^*_\tau);
\ J^{\mu }_{(s)}{}_{,\mu }=0.
\]
Here  $\gamma^*_\mu=s_{,\mu}s^{-1}$ is
the $so_3$- or $so(1,3)$-valued current (right-invariant).
In a slightly more detailed notations, with
$\gamma^*_{ij\mu }=s_{ik,\mu}s_{jk}$,
the topological current reads
\begin{equation}\label{tcs}
J^\mu _{(s)}=a\varepsilon^{\mu\nu\varepsilon\tau }
 \gamma^*_{ij\nu }\gamma ^*_{jk\varepsilon }\gamma^*_{ki\tau }.
\end{equation}
 The constant $a=\frac1{96\pi^2}$ is defined to
make topological charge $Q_{(s)}$ integer:
\[ Q_{(s)}=\int\! J_{(s)}^{0}\,d^{\,3} x \in Z=\pi _3(SO_3). \]
This charge has concern with (half of) the degree of the map
from the space (or Cauchy surface) ${\bf R}^3 $ to
the (sub)group $SO_3$ (which volume is $\pi^2$).

The topological current $J^{\mu }$ for a general frame field
$h^{a}{}_{\mu }(x^{\nu })$
should coincide with Eq.~(\ref{tcs}) when the metric is trivial,
$g_{\mu\nu}=\eta_{\mu\nu}$, and $h^{a}{}_{\mu }\in SO(1,3) $.
Let us use the language of differential  forms in order to find
the closed 3-form $t$ which is dual to the topological current,
$*t = J_\mu dx^\mu$.

It is known that characteristic classes  \cite{du}
are trivial in
(pseudo)Riemannian space  with a frame structure
(parallelizable manifold), so the class $c_{2}$
(which is the scalar-valued 4-form) is an
exact form:
\begin{equation}\label{c2}
c_2\equiv \tr(\Omega \wedge\Omega )= d u .
\end{equation}
One can use either
$so(1,3)$-valued connection (1-form $\gamma= \gamma_\mu dx^\mu$),
\begin{equation}\label{gamma}
 \gamma _{ab\mu }=
h_{b}{}^{\lambda }h_{a\lambda ;\nu }
=\gamma _{[ab]\mu },
\end{equation}
 or $gl_4$-valued connection $\tilde{\gamma}_\tau dx^\tau$
 with Christoffel symbol,
\[ \tilde{\gamma}^\mu{}_{\nu\tau }= -\Gamma^\mu _{\nu\tau }
=-{\textstyle{\frac{1}{2}}}g^{\mu\rho}
(g_{\rho\nu,\tau}+ g_{\rho\tau,\nu} -g_{\nu\tau,\rho}),
\]
that to write the (Riemann) curvature 2-form
$\Omega\equiv d\gamma +\gamma \wedge\gamma $
(and $\tilde{\Omega}$):
\begin{eqnarray}
&&\Omega_{ab\varepsilon\tau}=R_{ab\varepsilon\tau}=
2h^{b}{}_{\lambda}h^{a}{}_{\lambda ;[\varepsilon;\tau ]}
=\gamma_{ab\varepsilon,\tau }
+\gamma_{ac\varepsilon}
\gamma_{cb\tau}- (\varepsilon\tau ),
\nonumber\\
 &&\tilde{\Omega}^\mu{}_{\nu \varepsilon\tau } =
 R^\mu{}_{\nu \varepsilon\tau }=
-\Gamma ^\mu _{\nu \varepsilon ,\tau }+
\Gamma ^\mu _{\lambda \varepsilon }
\Gamma ^\lambda _{\nu \tau } - (\varepsilon \tau ).
\end{eqnarray}
Also, the (scalar-valued) 3-form $u$
can be written in two similar ways -- with and without
`tilde':
\begin{eqnarray}\label{uu}
 \tilde{u}= \tr(\tilde{\gamma}
  \wedge \tilde{\Omega}
 - {\textstyle{\frac{1}{3}}}
 \tilde{\gamma}\wedge\tilde{\gamma}
  \wedge\tilde{\gamma}), \
  u= \tr(\gamma \wedge \Omega-\cdots).
\end{eqnarray}
Using the Bianchi identity,
$d\Omega\equiv \Omega\wedge\gamma -\gamma\wedge\Omega$, and
\[\tr(\gamma \wedge
\gamma \wedge \gamma \wedge\gamma )\equiv0, \] one can verify that
$d u =d \tilde{u}=c_2$, see Eq.~(\ref{c2}). Therefore, we have the
closed 3-form $t\propto \tilde{u}-u$, $dt=0$, and can write the
topological current ($h=\det h^a{}_\mu=\sqrt{-g};
\, (hJ^\mu)_{,\mu}=0$):
\begin{equation}\label{tc}
 h J^\mu =
a\varepsilon ^{\mu \nu \varepsilon \tau }
\left(\Gamma^\alpha_{\beta\nu}
(3 \Gamma ^\beta _{\alpha \varepsilon, \tau }
-2\Gamma^\beta_{\gamma\varepsilon}\Gamma^\gamma_{\alpha \tau })
-\gamma _{ab\nu }
({\textstyle{\frac{3}{2}}} R_{ba\varepsilon \tau }
-\gamma_{bc\varepsilon}\gamma_{ca\tau })\right).
\end{equation}
Although $J^\mu$ is not of covariant view, it is
clearly appropriate for determination of topological
charge
\[ Q=\int\limits_V \! hJ^{0}\,d^3x \in \Pi=Z,
 \]
at least when the Cauchy surface is covered by one map.

At last, let's recall the Weitzenb\"{o}ck connection
from Eq.~(\ref{weitzen}),
 $\hat{\G}$, which may also be used as a
left-invariant (with respect to global Lorentz transformations)
1-form with zero curvature-form:
\[d\hat{\G}=-\hat{\G}\wedge \hat{\G} \ \
(\hat{\G}^\m_{\n\l}dx^\l=h_a{}^\m h^a{}_{\n,\l}dx^\l). \]
  So, the topological
current can also be  written in the next simple forms:
\begin{equation}\label{jw}
hJ^\mu =a\varepsilon^{\mu\nu\varepsilon\tau }
\hat{\G}^\a_{\b\nu}\hat{\G}^\b_{\g\varepsilon}\hat{\G}^\g_{\a\tau}
=  a\varepsilon^{\mu\nu\varepsilon\tau }
\hat{\g}^a{}_{b\nu}\hat{\g}^b{}_{c\varepsilon}\hat{\g}^c{}_{a\tau},
\end{equation}
where ($\hat{\g}$ and $\hat{\G}$ are not tensors)
\[\hat{\g}^a{}_{b\nu}=h^a{}_{\m,\n} h_b{}^\m=
h^a{}_\a \hat{\G}^\a_{\b\nu} h_b{}^\b \, .\]
\subsection{Spherical symmetry (in 4$D$); $SO_3$-quasi-charge 1-form}
The $SO_3$-symmetric frame field  can be
generally written as follows \cite{ei-ss,zh4}:
\begin{equation}  \label{spsy}
 h^{a}{}_{\mu }(t,x^i)=
 \pmatrix{a&bn_{i}  \cr
 cn_{i} & en_{i}n_{j}+d\Delta _{ij} +f\varepsilon_{ijk}n_k};
\ \ i,j=(1,2,3), \ n_i=\frac{x^i}{r}.
\end{equation}
 Here $a,\ldots,f$ are functions of time, $t=x^0$, and radius
 $r$, $\Delta_{ij}=\delta_{ij}-n_i n_j, \ r^2=x^i x^i$.
 As functions of radius, $a,e,d$ are even, while the others are odd;
 there are the boundary conditions: $e=d$ at $r=0$,
 and $h^a{}_\mu\to \delta^a_\mu$ as $r\to \infty$. Note that we
 use $i,j$ for space indexes of both Latin and Greek origin;
 this origin should be derived from the (a) context.

 It is easy to check that regarding $(t,r)$-transformation
 we have the next set of 2$D$-covariants: two co-vectors,
 $(a,b)$ and $(c,e)$, scalar $s=r\,d$, and pseudo-scalar $p=rf$.
 At points with $n_i=(1,0,0)$ we have the block matrix
\[  h^{a}{}_{\mu }=
\left(%
\begin{array}{cc} A & 0 \\ 0 & B \\
\end{array}%
\right); \ \ A=\left(%
\begin{array}{cc} a & b \\ c & e \\
\end{array}%
\right), \ B=\left(%
\begin{array}{cc} d & f \\ -f & d \\
\end{array}%
\right) =  \sqrt{d^2+f^2}\left(%
\begin{array}{cc}
  \cos\varphi & \sin\varphi \\
  -\sin\varphi & \cos\varphi \\
\end{array}%
\right).
\]
Here $\tan\varphi=f/d=p/s$. From this presentation one can
directly write the 2$D$-covariant 1-form of topological
quasi-charge, $q=q_{\tilde{\alpha}}d x^{\tilde{\alpha}}$ (we use
2$D$-indices $\bar{\alpha}=(t,r)$):
\begin{equation}\label{1form}
    q=\frac{d \varphi}{2\pi}=
    \frac{s p_{,\bar{\alpha}}-p s_{,\bar{\alpha}}}
    {s^2+p^2}\, \frac{d x^{\bar{\alpha}}}{2\pi}.
\end{equation}

The morphism
\[ Z=\Pi(SO_3;m=3) \to \Pi(m=3)=Z \]
is isomorphism. To prove this
statement and to check the constant $a$
in Eqs.~(\ref{tc}), (\ref{tcs})
 let us consider a `toy' static AP-space characterized
  by the next set of functions in (\ref{spsy}):
 \[ 
a=e=1, \ b=c=0,\ d=\cos\varphi(r),\ f=\sin\varphi(r),
\] 
with (trivial metric, $g_{\m\n}=\eta_{\m\n}$, and)
boundary conditions
\[\cos\varphi(0)=\cos\varphi(\infty)=1; \mbox{ \ hence }
\sin\varphi(0)=\sin\varphi(\infty)=0.\]
Substituting this frame field into Eq.~(\ref{gamma})
we obtain the none-zero
components of $\gamma_{ab\mu}$:
\[\gamma_{ijk}=\frac{CS}r \varepsilon_{ijk}
+(\varphi' -\frac{CS}r) \varepsilon_{ijl}n_l n_k
+\frac{1-C}r \left(
S(n_i\varepsilon_{jkl}-n_j \varepsilon_{ikl})n_l
-n_i\delta_{jk} + n_j\delta_{ik}\right) , \]
where the prime ($\,{}'$) denotes the derivative with
respect to radius, $C=\cos\varphi, \ S=\sin\varphi$.

At points with $n_i=(1,0,0)$ we have the next
 values (keep in mind that
$\gamma_{ab\mu}=-\gamma_{ba\mu}$):
\begin{eqnarray}
\nonumber
&& \gamma_{231}=\varphi' , \
\gamma_{232}=\gamma_{233}=0
=\gamma_{121}=\gamma_{131};  \\ \label{123}
  && \gamma_{123}=\gamma_{312}=S/r, \
\gamma_{212}=\gamma_{313}=(1-C)/r.
\end{eqnarray}
At last we are ready to find from Eq.~(\ref{tc}), or (\ref{tcs}),
the density of topological charge using (\ref{123}):
\begin{eqnarray}
\nonumber
hJ^0\!&=&a\varepsilon^{ijk}
 \gamma_{abi}\gamma_{bcj}\gamma_{cak}=
6a \varepsilon^{ijk}\gamma_{23i}\gamma_{31j}\gamma_{12k}\\ &{=}&
6a \gamma_{231}(\gamma_{312}\gamma_{123}-\gamma_{313}\gamma_{122})
  =12a\varphi' (1-\cos\varphi )/r^2.
\end{eqnarray}
Taking a `unit kink', with
$\Delta\varphi=\varphi(\infty)-\varphi(0)=2\pi$,
see the boundary conditions,
we should obtain the unit topological charge:
\[Q=4\pi\int_0^\infty\!\! hJ^0 r^2 dr=96a\pi^2=1; \mbox{\ hence }
a=\frac1{96\pi^2}. \]

\subsection{Cylindrical symmetry; $SO_2$-quasi-charge}
In this section we are going to show that $SO_2$-quasi-charge
(which is equal to topological charge as well) can be
completely attributed (in a covariant manner)
to the axis of cylindrical symmetry
(that is, the line of stationary points on a Cauchy surface).

Let us consider the Killing vector, $^{(3)}\xi$, corresponding to
cylindrical symmetry with respect to axis $x^3$; that is, in
 `natural Cartesian' coordinates (where $\xi^\m{}_{,\n}={}$const),
 \[ ^{(3)}\xi={}^{(3)}\xi^\m\, \pa_\m = x^1\pa_2-x^2\pa_1\, .\]

 In general, $\xi$ is a Killing vector if some combination of
 infinitesimal ($\eps\ll1$)
  coordinate transformation and global (Lorentz) rotation,
 \[ x^\m \to \tilde{x}^\m=x^\m+\eps\,\xi^\m, \
 s^a{}_b=\d\,^a_b+\eps\, m^a{}_b\in SO(1,3)\cap SO_3\, ,\]
 does not change frame field:
 \[\tilde{h}^a{}_\m(\tilde{x})=(\d\,^a_b+\eps\, m^a{}_b)
 h^b{}_\n(x)(\d\,^\n_\m-\eps\, \xi^\n{}_{,\m})=h^a{}_\m(\tilde{x}).\]
 This condition leads to the next equation
 ($ m^a{}_b={\rm const}$)
\begin{equation}\label{xi}
 h^a{}_{\m,\n}\xi^\n=m^a{}_b h^b{}_\m
 - h^a{}_\n\xi^\n{}_{,\m} \,,
 \end{equation}
which can easily be written in manifestly covariant form: 
$(h^a{}_\n\xi^\n)_{;\m}+\L^a{}_{\m\n}\xi^\n=m^a{}_b h^b{}_\m\,$. 
(Its symmetric part gives usual equation of GR:
$\xi_{(\m;\n)}=0$).
The set of {\em stationary points}, where
\begin{equation}\label{sp}
\xi^\m=0, \ \xi^\m{}_{,\n}= h_a{}^\m m^a{}_b h^b{}_\n\,,
\end{equation}
will be denoted  $\Xi_0$.

The matrix $m$ of the $x^3$-axial symmetry looks as follows:
\[ m^a{}_b=
{}^{(3)}m^a{}_b= -\ve^{03ab} ,\]
and in 'natural coordinates'
${}^{(3)}\xi^\m{}_{,\nu}={}^{(3)}m^\m{}_\n$.
(It is assumed that at spatial infinity $h^a{}_\mu=\d^a_\m$.)

 Let us introduce
 cylindrical coordinates, $y^M$,
 suitable to enumerate
orbits of the symmetry:\footnote{
In new coordinates Killing vector is constant,
$\xi^M=(0,0,0,1)$, except when $\rho=0$.}
\[ y^M=(t,z,\rho,\phi); \ x^0=t, \ x^1=\rho\cos\phi,
 \  x^2=\rho \sin\phi, \ x^3=z .  \]
We want to use the Eq.~(\ref{jw}) for topological current,
keeping Cartesian indices for frame matrices but switching
to cylindrical ones in derivatives.
One can show (using (\ref{xi}) and its prolongation)
 that this current does not depend on $\phi$:
\[ (hJ^\m)_{,\phi}=0=(hJ^\m)_{,\n}\xi^\nu=
(hJ^\m)_{,\n}\frac{\pa x^\nu }{\pa \phi}  \,.\]
In other words, in matrix notations (with bold letters),
 Eq.~(\ref{xi}) leads to
\[ { \emph{\textbf{h}}}_{,\phi}=\emph{\textbf{mh}}-
\emph{\textbf{hm}}\, \
\mbox{ (hence }
\ \tr(\emph{\textbf{h}}^{-1}\emph{\textbf{h}}_{,\phi})
=0=(\det\emph{\textbf{h}})_{,\phi})\]
\[ \mbox{and } \
\emph{\textbf{{h}}}(\phi)=e^{\emph{\textbf{{m}}}(\phi-\phi_0)}
{\emph{\textbf{ h}}}(\phi_0)e^{-\emph{\textbf{{m}}}
(\phi-\phi_0)}\,.\]
Therefore,
substituting above equations in Eq.~(\ref{jw})
and making integration over $\phi$, we can obtain
the topological quasi-current,
$hJ^X_{(SO2)} =\int  hJ^X \rho\, d\phi$,
 for $SO_2$-symmetry
(in mixed coordinates; $X,Y=(t,\rho,z)$; $m^\m{}_\n$
is constant and not a tensor):
\begin{equation}\label{j2so2}
hJ^X_{(SO2)} = 6\pi a
\varepsilon^{X YZ \phi} (m^a{}_b
\hat{\g}^b{}_{c Y}\hat{\g}^c{}_{a Z}-
m^\m{}_\n \hat{\G}^\n_{\l Y}\hat{\G}^\l_{\m Z})
= 6\pi a \varepsilon^{X YZ \phi} (m^a{}_b
\hat{\g}^b{}_{a Z} + m^\m{}_\n \hat{\G}^\n_{\m Z})_{,Y}\,,
\end{equation}
where $\hat{\g}^a{}_{bX}=h^a{}_{\m,X} h_b{}^\m, \
\hat{\G}^\m_{\n X}=h_a{}^\m  h^a{}_{\n,X}  $.
This quasi-current is `trivial' (ie, corresponds to
an exact form); making integration over radius $\rho$
one can obtain:
\begin{equation}\label{jso2}
Q  =\int\! hJ^{0}dV=Q_{(SO2)}=
\int\!\!\int\! hJ^{\,t}_{(SO2)}d\rho\, d z =12\pi a
\int^\infty_{-\infty}\!\!
  \left. m^a{}_b
 h^a{}_{\m,z} h_b{}^\m dz\right|_{\rho=0}
\, .
\end{equation}
It is taken into account that
$\left.m^\m{}_\n=
h_a{}^\m m^a{}_b h^b{}_\n\right|_{\Xi_0}$,
see Eq.~(\ref{sp}).

More over one can easily check
(differentiating Eq.~(\ref{sp}) with proper contracting)
that at stationary points it is valid:
$\left.
h_a{}^\m {}_{,\m}\right|_{\Xi_0}=0$. Then, taking
any space-like line, $\zeta$, placed on $\Xi_0$:
\[ \zeta: \ {\bf R} \to \Xi_0 \subset  {\bf R}^{1+3},
\ g_{\m\n}\frac{d x^\m(\zeta)}{d \zeta}
\frac{d x^\n(\zeta)}{d \zeta} >0\, ,\]
one can can arrive at the next covariant line
density of topological (quasi)charge:
\begin{equation}\label{jcso2}
Q   =12\pi a
\int^\infty_{-\infty}\!\!
 m^a{}_b
 (h^a{}_{\m,\nu}-h^a{}_{\n,\m}) h_b{}^\m
 \frac{d x^\m(\zeta)}{d \zeta}\,d\zeta
\, .
\end{equation}
This equation, being covariant
($h^a{}_{\m,\nu}-h^a{}_{\n,\m}$ is the tensor, $\L$),
should be valid in any coordinates, not only
`natural Cartesian'.
This is an interesting result -- in view of possible
generalization for the next $D$.

At last, taking the $SO_3$-symmetrical $h$-field of
the previous subsection, one can easily check
(components in (\ref{123}) can be used after
permutation \{3,1,2\}) that Eq.~(\ref{jso2}) gives:
\[ Q   =24\pi a
\left.\int^\infty_{-\infty}\!
\frac{d\vphi}{dr}\,dx^3\right|_{x^1=x^2=0}
=96 \pi^2 a =1\, .
 \]
\section{`Left' and `right'
topological (quasi-)charges in 5$D$}
In this section $m=4$.
We need for two topological currents $J^\mu_{(l)}$ and
$J^\mu_{(r)}$, `left' and `right', which should replace each
other under $P$-inversion, or  $C$-operation (see below).
Now we are going to use the quaternion representation of the $SO_4$
group
that to write quaternion differential forms of topological
charges (`left' and `right')  and quasi-charges.

\subsection{Quaternion representation of $SO_4$}
With quaternion units denoted as ${\bf i}_k$
(that is,
${\bf i}_1={\bf i},\ {\bf i}_2={\bf j},\ {\bf i}_3={\bf k}$),
the quaternion multiplication is defined by the rule:
\begin{equation}\label{ijk}
 {\bf i}_j\, {\bf i}_k = - \d_{jk} + \ve_{jkl}\,{\bf i}_l
 , \ j,k,l=1,2,3.
\end{equation}
Any quaternion
${\bf x}\in{\bf H}={\bf R}^{4}$ has real and imaginary parts,
and the module $|{\bf x}|$ (absolute value):
\[ {\bf x}= 
x_{4}+x_k {\bf i}_k=\Re(\textbf{x})+\Im(\textbf{x}),
 \ |{\bf x}|^2={\bf x}{\bf \bar{x}}
=x_4^2+ x_k x_k\, , \]
where ${\bf \bar{x}}=x_4-x_k\, {\bf i}_k$ -- conjugate quaternion.
The set of quaternions  with absolute value one,
$\textbf{H}_1=\{\textbf{f},\ |\textbf{f}|=1\}$, forms a
group under quaternion multiplication,
and $\textbf{H}_1\cong SU_2=S^3$.

Any $s \in SO_{4}$
can be represented as a pair of such unit length quaternions
\cite{du},
$({\bf f},\,{\bf g})\in
S^{3}_{(l)}\times S^{3}_{(r)},\
|{\bf f}|=|{\bf g}|=1 $:
\[ x^*=sx \ \Leftrightarrow \ {\bf x}^*=
{\bf f\,x\,g}^{-1}={\bf f\,x\,\bar{g}} \,
; \ \ |{\bf x}|=|{\bf x^*}|. \]

The pairs (\textbf{f},\,\textbf{g}) and
(--\textbf{f},\,--\textbf{g}) correspond to
the same rotation $s$, that is,
\[SO_{4}=S^{3}_{(l)}\times S^{3}_{(r)}/\pm \, .\]
We indicate these quaternion spheres
as {\sl left\/} and {\sl right},
 because under space inversion $P_3\{1,2,3\}$,
and under the $C-$operation
\footnote{This letter (C) is used that some allusion
to the charge conjugation is to arise.}, which
 inverts the 4-th coordinate, $C=P_1\{4\}$,
pair elements replace one another:
$ C: \, ({\bf f},{\bf g})\to
({\bf g},{\bf f})$;
({\bf f},\,{\bf f})\,$\in SO_{3}$.

The rotations on angle $\phi $
of coordinates
(1) $x_2,x_3$, \  (2) $x_{4},x_1$ have the next quaternion
form (they commute): 
\begin{equation}
\label{aa} (1) \
({\bf a},{\bf a})(\phi ),\
(2)  \ ({\bf a},{\bf \bar{a}})(\phi)
;  \ {\bf a}(\phi)=
\cos\frac\phi 2+{\bf i}\,\sin\frac\phi 2
=e^{{\bf i}\,\phi/2}.
\end{equation}
Note that the symmetry condition (\ref{gsi}) also splits
into two parts:
\begin{equation}
\label{ab}
 {\textbf{f}\/(\textbf{a}\/\textbf{x}\/\textbf{b}}
 {}^{-1})=\textbf{a}\/\textbf{f}(\textbf{x})\/\textbf{a} {}^{-1},
 \ \textbf{g}\/(\textbf{a}\/\textbf{x}\/\textbf{b}{}^{-1})
 =\textbf{b}\/\textbf{g}(\textbf{x})\/\textbf{b}{}^{-1} \ \forall
(\textbf{a},\/\textbf{b})\/ \in G \subset SO_{4}.
\end{equation}

Following the Fjodorov's parametrization of $SO_4-$group \cite{fjo}
one can define on $so_4-$algebra  the `left' and `right' (or
selfdual and anti-selfdual) `imaginary units'
\begin{equation}\label{iu-s}
    M^{(\pm i)}=L^{(i)}\pm  K^{(i)}, \
L^{(i)}_{ab}=     -\ve_{abi4}, \
    K^{(i)}_{ab}=\d_{a4}\d_{bi}- \d_{b4}\d_{ai}.
\end{equation}
For example (zeroes are not shown in these matrices),
\[M^{(+1)}=\left(%
\begin{array}{cccc}
   \ & \  & \ & -1 \\
  \ & \ & -1 & \ \\
  \ & 1 & \ & \ \\
  1 & \ & \ & \ \\
\end{array}%
\right), \ M^{(+2)}=\left(%
\begin{array}{cccc}
   \ & \  & 1 &   \\
  \ & \ &   & -1 \\
  -1 &   & \ & \ \\
    & 1 & \ & \ \\
\end{array}%
\right), \ M^{(+3)}=\left(%
\begin{array}{cccc}
   \ & -1  &   &   \\
  1 & \ &   &   \\
    &   & \ & \ -1 \\
    &   & 1 & \ \\
\end{array}%
\right). \]
It is easy to check that
\[ [M^{(+i)},M^{(-j)}]=0, \ \tr(M^{(+i)}\,M^{(-j)})=0,
\] \[
M^{(\pm i)}M^{(\pm j)}=-\d_{ij} E +\ve_{ijk} M^{(\pm k)},
\ \tr(M^{(\pm i)}\,M^{(\pm j)})=-4\d_{ij},\ \]
where $E=1^4$ is the identity matrix; compare the last equation
with the Eq.~(\ref{ijk}).

The separation of any matrix $s\in SO_4$ into selfdual and
anti-selfdual parts looks as follows:
\[s=s^{(+)}\,s^{(-)}=s^{(-)}\,s^{(+)}, \
s^{(+)}=f_4 E+ f_i M^{(+i)},\ s^{(-)}=g_4 E+ g_i M^{(-i)}. \]
It is evident that $s\, s^{tr} = 1$
($tr$ indicates the transposed matrix), and that
\[f_4 g_4 = \fr14 \tr(s), \
g_4 f_i = -\fr14 \tr(M^{(+i)}s),
\ f_4 g_i = -\fr14 \tr(M^{(-i)}s)\,.\]

\subsection{Example of $SO_2$-symmetric quaternion field}
Let's consider an example of $SO_2\{2,3\}-$symmetric \textbf{f}--field
 configuration (\textbf{g}=1), which carries both charge and
$SO_2$-quasi-charge (left, of course),
 $\textbf{f}(\textbf{x}){:} \
 {\bf H}={\bf R}^4 \to {\bf H}_1; \ \textbf{f}(\infty)=1$.
The symmetry condition from (\ref{aa}), (\ref{ab}) reads
\begin{equation}\label{f-so2}
 \textbf{f}(e^{{\bf i}\,\phi/2}\textbf{x}e^{-{\bf i}\,\phi/2})
   =e^{{\bf i}\,\phi/2} \textbf{f}(\textbf{x}) e^{-{\bf i}\,\phi/2}.
\end{equation}
We'll switch to `double-axial' coordinates:
$\textbf{x}=a e^{{\bf i}\,\vphi}+ be^{{\bf i}\,\psi}{\bf j}$.
Let us use  imaginary quaternions $\textbf{q} $ as stereogrphic
coordinates on $ {\bf H}_1$, and take symmetrical field
$\textbf{q}(\textbf{x}) $ consistent with Eq.~(\ref{f-so2}):
\begin{equation}\label{fq(x)}
\textbf{q}(\textbf{x})=\textbf{x\,i}\,\bar{\textbf{x}}+\textbf{i}
=-\bar{\textbf{q}}, \ \textbf{f}(\textbf{x})=
-\frac{1+ \textbf{q} }{1- \textbf{q}}=
1-\frac{2}{1- \textbf{q}}.
\end{equation}
It is easy to find the `center of quasi-soliton'
(1-submanifold, $S^1$)
\[ S^1=\textbf{f}^{-1}(-1)=\textbf{q}^{-1}(0)
=\{a=0,\ b=1\} =
\{\textbf{x}_0,\ \textbf{x}_0=e^{{\bf i}\,\psi}{\bf j}\} \]
and the `vector equipment' on this circle:
\[ d \textbf{x}|_{\textbf{x}_0}\left.
=da\, e^{{\bf i}\,\vphi}+ (db +\textbf{i}\, d\psi)
e^{{\bf i}\,\psi}{\bf j}, \
\fr1 4  d \textbf{f}\right|_{\textbf{x}_0}=
{\bf i}db-{\bf k}\,
e^{{\bf i}\,(\vphi+\psi)} da\, ; \]
\textbf{i}-vector all time looks along the radius $b$
(parallel translation along the circle $S^1$; this is a
`trivial`, or `flavor'-vector). Two others
('phase'-vectors) make $2\pi-$rotation along the circle.

In fact, the field (\ref{fq(x)}) has also symmetry
$SO_2\{1,4\}$, and this feature restricts possible
directions of `flavor'-vector (two `flavors' are
possible, $\pm$; the $P_2\{1,4\}-$symmetry gives
the same effect).
The other interesting observation is that the
equipped circle
 can be located also at the stationary
points of $SO_2-$symmetry (this increases
the number of `flavors').

The left 3-form of topological quasi-charge can be written
as follows:
\[ q_{(l)}=\fr1{12 \pi^2} \Re(\textbf{r}\wedge \textbf{r}
\wedge \textbf{r})=
\fr1{12 \pi^2} \Re(\textbf{l}\wedge \textbf{l}
\wedge \textbf{l})
,\]
where $\textbf{r}$ (resp. $\textbf{l}$)
 is right- (left-) invariant quaternion-valued 1-form
\[ \textbf{r}={\bf f}_{,A}\,\bar{\bf f}\, dA, \
\textbf{l}=\bar{\bf f}\,{\bf f}_{,A}\, dA, \
\, A,B = \{r_1,r_2,\vphi_2\}; \
d\textbf{r}=\textbf{r}\wedge\textbf{r}, \
d\textbf{l}=-\textbf{l}\wedge\textbf{l}  . \]
(One can use Pauli matrices instead of  quaternion units
replacing $\Re()$ with $\fr12\tr()$. The left 2-form of
$SO_2\times SO_2-$quasi-charge can be written as well,
and the right forms as well.) 

\subsection{Hopf mapping and quaternion forms} %
Imaginary quaternions ($\textbf{J}\subset \textbf{H}$) of unit length,
${\bf n}$,
form 2-sphere:
$${\bf n}=n_j {\bf i}_j \in {\bf J}\cap {\bf H}_1=S^2 .$$
The Hopf (and `anti-Hopf') map takes $S^3={\bf H}_1$ into  $S^2$:
\begin{equation}\label{hopf}
 H_{\bf n}: \,{\bf f}\mapsto{\bf w}={\bf f}\,{\bf n}\,{\bf\bar{f}}
 \ (\bar{H}_{\bf n}: \,{\bf f} \mapsto {\bf w}
={\bf \bar{f}}\,{\bf n}\,{\bf f}).
\end{equation}
It is evident that
$ \bar{{\bf w}}=-{\bf w}, \ {\bf w}^2=-1,\
 {\bf w}\, d{\bf w}=-d{\bf w}\,{\bf w}\,. $
\subsubsection{Map $\textbf{R}^{2+1}\to S^2$
and charges from $\pi_2(S^2)=Z$.}
 The unnormalized 2-form of topological charge (relating to
`volume' (surface) form on $S^2$) is
\[ \underline{t}^{(2)}=  
\Re({\bf w}\, d{\bf w}\wedge d{\bf w})\,; \
d\,\underline{t}^{(2)} = \Re(d{\bf w}\wedge d{\bf w}\wedge d{\bf w}) =
\mp\, \Re({\bf w}\,d{\bf w}\wedge d{\bf w}\wedge d{\bf w}\,{\bf w})
=0.\]
We use the same letter for `pull-back' forms on
$\textbf{R}^{2+1}$, say, \, $d{\bf w}={\bf w}_{,\m} d x^\m$.

The normalizing coefficient
 is $1/(8\pi)$: $4\pi$ is the surface of $S^2$
and factor 2 is due to $\ve$-permutation in 2-form.
That to prove this
one can use the stereographic projection on the `complex
plane', $\textbf{C}=\{\textbf{z} = x + \textbf{i} y\}$:
\[ \textbf{w}= \textbf{i} \frac{1+\textbf{kz}}{1-\textbf{kz}}
=\textbf{i}\frac{2}{1-\textbf{kz}}-\textbf{i}=
(1-\textbf{kz})\textbf{i}(1-\textbf{kz})^{-1}=
\frac{1-\textbf{kz}}{1+\textbf{kz}}\textbf{i} ; \ \,
\]
\[d\textbf{w}=2\textbf{i}(1-\textbf{kz})^{-1}
\textbf{k} d\textbf{z}(1-\textbf{kz})^{-1}; \
\textbf{k} d\textbf{z}  =  d\bar{\textbf{z}}\, \textbf{k} \,; \]
\[
t^{(2)}=\frac1  {8\pi}\Re({\bf w}\, d{\bf w}\wedge d{\bf w})=
-\frac{\Im(d\textbf{z}\wedge d\bar{\textbf{z}})}%
{2\pi(1+||^2)^2}=
\frac{dx\, dy}{\pi(1+x^2+y^2)^2}.
\]
The other way is to use first the symmetry of
 axisymmetric field
$\textbf{w}(\textbf{z})$,
 see (\ref{f-so2}), which
complies with
the following condition (in axial coordinates,
$ \textbf{z}{=}\rho e^{\textbf{i} \vphi}$;
see the field considered above):
\begin{equation}\label{d-so2}
\textbf{w}(\rho e^{\textbf{i} \vphi})=
e^{-\textbf{i} \vphi/2}\textbf{w}(\rho )e^{\textbf{i} \vphi/2},
\ \, 2\textbf{w}_{,\vphi}
 =\textbf{w}\textbf{i}-\textbf{i}\textbf{w}.
\end{equation}
Substituting this into 2-form and making integration
one can check that ($\textbf{w}(0)=\textbf{i},\
 \textbf{w}(\infty)=-\textbf{i} $)
\[ \int_{\textbf{z} }\underline{t}^{(2)}=
\int_{\rho,\phi} \Re({\bf w}\,
d{\bf w}(\textbf{w}\textbf{i}-\textbf{i}\textbf{w}))
\wedge d\vphi
=4\pi\int_0^\infty \Re(\textbf{i}\textbf{w}_{,\rho})
d \rho = 8\pi\,. \]
\subsubsection{Maps $\textbf{R}^{3+1}\to S^3, S^2$ and charges from
$ \pi_3(S^3)=\pi_3(S^2)=Z$.}
\subsubsection{Maps $\textbf{R}^{4+1}\to S^3, S^2$ and charges from
$ \pi_4(S^3)=\pi_4(S^2)=Z_2$.}

(To be continued; the transition $s \to h$.)

\subsection{Quasi-charges and their morphisms in 5$D$}
First of all one should consider those symmetries, $G$, which
are contained in the symmetries of cosmological background solution:
 \be
  G \subset G_0 = (O_3\times P_4)\cap SO_4\, .
  \ee
  Here $P_4$ is the inversion of all space-like coordinates; $G_0 $
  is the point symmetry of `wave-guide' in co-moving coordinate
  system.
  It is assumed that weak (or not so weak)
  stochastic waves (which have also
  ponderable $f$-component) can decrease symmetry of large-scale
  solution (which is $O_4$, or in co-moving system results
  in $O_3\times P_4$) down to $G_0 $.

  If $G\subset G_0$, the QC-group has two isomorphous parts,
  left and right:
  $$\Pi(G)=\Pi_{(l)}(G)+\Pi_{(r)}(G) . $$

  The following Table describes quasi-charge groups and
  morphisms (for details see \cite{the}); here
  the 4-th coordinate is the
  extra dimension which looks along the radius of $G_0 $-symmetric
  expanding cosmological background.  \\[1mm]

{\bf Table. } QC-groups  $\Pi_{(l)}(G)$ and their morphisms
to the preceding group; $G\subset G_0$.
\\[2mm] \hspace*{\parindent}
\begin{tabular}{c|c|c}
 $G $ &$ \Pi_l(G)\to\Pi_l(G^*) $
&$ \mbox{`label'}$ \\ \hline\hline
$1 $&$ Z_2 $&  \\[1mm] \hline
$SO\{1,2\} $&$ Z_{(e)} \stackrel{e}{\to}Z_2 $&$ e $ \\[1mm] \hline
$SO\{1,2\}\times{}P\{3,4\} $&$ Z_{(\nu)}+Z_{(H)}
\stackrel{i,m2}{\to}Z_{(e)} $&$ \nu^0;\ H^0\to e+e \ \
$ \\[1mm] 
$SO\{1,2\}\times{}P\{2,3\} $&$ Z_{(W)}
\stackrel{0}{\to}Z_{(e)} $&$ W\to e+\nu^{0}$ \\[1mm]
$SO\{1,2\}\times{}P\{2,4\} $&$ Z_{(Z)}
\stackrel{0}{\to}Z_{(e)} $&$ Z^0\to e+e$ \\[1mm]\hline
$SO\{1,2\}\times{}P\{3,4\}\times{} $&$ Z_{(\gamma)}
\stackrel{0}{\to}Z_{(H)} $&$ \gamma^0\to H^{0}+H^{0}$
 \\[0mm]
$ \times{}P\{2,3\} $&$ \ \ \ %
\stackrel{0}{\to} Z_{(W)}$ &$ \ \ \ \to W+W$
 \end{tabular}  \\[1mm]
It seems that `quasi-particles', which symmetry includes $P_4$,
should be true neutral (neutrinos, Higgs particles, photon).

  One can assume further that an hadron bag is a specific place
  where $G_0-$symmetry does not work, and the bag's symmetry
  is isomorphous to $O_4$ (or $SO_4$).
  This assumption can lead to
  another classification of quasi-solitons (some doubling the
  above scheme), where self-dual and anti-self-dual one-parameter
  groups (e.g., (a$^2$,1) and (1,a$^2$), see Eq.~(\ref{aa}) for `a')
  take place of $SO_2-$group.
  The total set of quasi-particle parameters
  (parameters of equipped 1-manifold (loop)  plus
  (or `times')
parameters of group) for (anti)self-dual groups
(or for $SO_2\subset O_4$),
  real Grassman manifold $G(4,2)$ times $RP^2$,
  is larger than the analogous set for
  groups $SO_2\subset G_0$,
  which is just $O_3$ times $G(3,1)=RP^2$ .
If the number of `flavor'-parameters (which are not degenerate and
have some preferable particular values)
is the same as in the case of
`white' quasi-particles, the remaining parameters (degenerate, or
`phase') can give some room for `color' (in addition to spin).

   So, perhaps one might think about `color neutrinos'
  (in the context of pomeron, and baryon spin puzzle),
`color W, Z, and Higgs' (another context -- say, $B$-mesons),
and so on.

Note that in our picture the very notion of quasi-particle depends on
the background symmetry. On the other hand, large clusters of
quasi-particles (matter) can disturb the background, i.e. the form
(and thickness) of waveguide; and waves of such
small disturbances (with
wavelength  larger than the thickness $L$, sure)
can be generated as well (but these waves do carry
no (quasi)charges, that is, are {\sl not quantized}).
Phenomenology of {\sl gravity} phenomena can arise
as an inner (i.e., (1{+}3)$D$), stable (i.e., Lagrangian)
phenomenology describing
 the form (curvature and thickness)
  of cosmological waveguide filled with quasi-particles
(and noise).

\section{Conclusion (need for a crucial experiment)}
This section is arranged as a letter to a professor,
whose Lab
is well equipped for the Bell-type experiments. These
experiments\cite{aspect,weihs}
 (for a review see \cite{tit})
meet great interest and
continuing discussions\cite{santos,khr,gis}.
(Recently, Fellows has proposed an interesting classic
model\cite{fel}, arranged as a circus knife-throwing
demonstration, where the Bell inequality is violated;
however, this model does not meet
the dichotomy requirement; in other words,
 the portion of `half-empty' events, (+,\,0),
(0,\,--), and so on (only one knife of two reaches its target
at `plus'- or `minus'-part), and, hence,
 the {\sl detectors' efficiency}
 does vary with the angle between the {\sl `polarizers'}.
Well, we believe that Quantum Mechanics gives
the excellent description,
but QM does not know what an {\sl objective reality} is
under description.)
So, the letter is as follows:\\[2mm]
Dear Prof NN,\\
today many laboratories (including yours) have sources of single
(heralded) photons, or entangled bi-photons. Moreover, at some
universities students can perform laboratory works
with single photons, having convinced on their own experience,
that light is quantized, and the classical description is
incorrect (the Grangier experiment)\cite{sites}.

I would like to suggest to your attention a minor modification
of the experiment with single photon interference, say, in a
Mach-Zehnder fiber interferometer with `long' (the fibers may be
rolled) enough arms, 3--5 km (that the time of flight of
`photon's halves' to be 10--20 $\mu$s), or, what may be even easier
for your Lab, -- with bi-photon non-local (also `long' arms --
to Alice and Bob) correlations of photon polarizations (the
Bell-type experiment).

The objective of this experiment is verification of
some naive-realistic model (`spaghetti-model', perhaps more
naive than realistic), relating to the origin (or possible
non-local 5$D$ ontology -- with still point-like 4$D$ form-factor)
of elementary particles, photons. According to this model, there
is one extra space dimension (the radial dimension in an
$O_4-$symmetric non-stationary `cosmological' solution of some
5$D$
field theory), and `particle' is a spaghetti-like non-linear
field configuration very elongated
(say,\footnote{
Some rough estimates are possible; the most naive is:
they say that the Pioneer's effect {\em turns on} at 10 AU, while
the antenna pattern (angle $\l/d$) is 0.01; multiplication
of these two gives 10$^{12}$ cm.} $L\sim10^{12}$ cm)
along this extra-direction
and carrying a topological (quasi)charge.

The new element in the proposed experiment is a fast-acting
shutter placed at the beginning of one of the interferometer's
arms (or at the beginning of the Bob's arm -- he should carry
out more work). The closing-opening time of the shutter should
be small enough, say 5--10 $\mu$s.
For example, if there is an air gap
in this arm (the length of gap may be variable that to draw the
interference figure), one can use a quickly rotating metal disk
with a set of holes (perhaps, of different diameter -- for
the  sake of comparison).

Both Quantum mechanics (no particle's ontology) and Bohmian
mechanics (wave-particle double ontology)\cite{nik}
 exclude any change in
the interference figure as a result of separating activity of
such a fast shutter (while the photon's `halves' are making their
ways to the place of a meeting), but I assume (that is,
my calculations show) that
\emph{a significant decrease of the interference
visibility can be observed}.\footnote{Measuring this visibility
as a function of interferometer's length and transparency of
fast shutter, one can arrive at some ideas about
the value of $L$.}

Certainly, the usage of electro-optical modulator
(in combination with polarizer -- for single photon case)
may turn
out to be preferable -- that to ensure the full `separation of
photon's halves', or full absence (impossibility of bypass
passage) of some `remnant photon's field', or any `links',
`navel-strings', linking the `halves': in this case only one of
the two will further carry topological quasi-charge (becoming a
`full photon'), while the other will dissolve in `zero-point
oscillations'. QM is everywhere (where we can see, of course),
 and, so, non-linear 5$D$-field
fluctuations, looking like spaghetti-anti-spaghetti loops,
should exist everywhere. (Perhaps, this omnipresence can be
related to the universality of `low-level heat death',
restricted by
the presence of topological quasi-solitons -- some as
 the 2$D$ computer experiment by Fermi, Pasta, and Ulam,
where the process of thermalization was restricted by
the existence of solitons.)

Having no doubts in your scientific responsibility (I was
reading your papers, including \ldots), I
 realize at the same time that strong enough motivation
(and some determination or grit)
should exist that to make any new experiment
(especially if this experiment is some questionable --
like the first measurement of proton's magnetic moment
(Otto Stern); the suggested experiment is much more
simple than, say, the single photon experiment performing
by Marshall et al\cite{marshall}).
That to explain somehow my own motivation (that it is not mere
my fantasy), there is the preceding paper
concerning topological charges, quasi-charges, and their
differential forms in a 5$D$ variant of
gravitation theory with Absolute Parallelism; 4$D$ (quasi)charges
are also considered for methodological purposes.
(Any your comments are welcome. Frankly speaking, I am
hoping not so much on `my motivation', as on your intuition and
deep interest in the puzzle of non-local correlations.)

AP, at least at the level of its symmetry,
seems to be able to cure the
gap
between the two branches of physics -- General Relativity
(with coordinate diffeomorphisms) and Quantum Mechanics
(with Lorentz invariance).\footnote{
Rovelli writes\cite{rov}:
{\it In spite of their empirical success, GR and QM offer
a schizophrenic and confused understanding
of the physical world}.}
Most people give all the rights of fundamentality (or primacy) to
quanta, and so, they try to quantize the gravity, and the
very space-time (probing loops, and strings, and branes;
see also the warning polemic by Schroer \cite{schroer}).
The other possibility is that quanta have a phenomenological
origin of a very specific kind
(relating to topological charges and quasi-charges).

With my best regards, \ldots


\small
\begin{thebibliography}{99} 
\bibitem{sa}T. Sauer,
 {\it Field equations in teleparallel spacetime:
Einstein's Fernparallelismus approach towards unified field
theory,\ }
%
\url{http://arxiv.org/abs/physics/0405142}

\bibitem{var}
 J.G. Vargas, {\it Geometrization of 
  Physics
 with Teleparallelism}, Found.\ Phys.\
{\bf 22}, 507 (1992).

\bibitem{itin}
 Y.\ Itin, {\it Energy-momentum current for coframe gravity},
   \href{http://arXiv.org/abs/gr-qc/0111036}%
{\tt arXiv.org/gr-qc/0111036}.

 \bibitem{agp}
  V.C. de Andrade, L.C.T. Guillen and J.G. Pereira,
 {\it Teleparallel gravity: an overview}, \\
 \url{http://arXiv.org/abs/gr-qc/0011087}.

\bibitem{vas}
 D. Vassiliev, {\it A teleparallel model for the neutrino},
  \url{http://arXiv.org/gr-qc/0604011}.

\bibitem{eima}
 A. Einstein and W. Mayer, Sitzungsber.\ preuss.\ Akad. Wiss.
 {\bf Kl}  257--265 (1931).

 \bibitem{ei-ss}
 A. Einstein and W. Mayer, Sitzungsber.\ preuss.\ Akad. Wiss.
 {\bf Kl}   110--120 (1930).

 \bibitem{uc}
A. Unzicker and T. Case, Translation of Einstein's attempt of a
unified field theory with teleparallelism (2005), 
\url{http://arxiv.org/abs/physics/0503046}.

 \bibitem{zh3}I.L. Zhogin, J. Soviet Physics {\bf 34},
 105 (1992); \
  \url{http://arXiv.org/abs/gr-qc/0203008}. 

 \bibitem{zh4} I.L. Zhogin, J. Soviet Physics {\bf 34},
 781 (1992); \
  \url{http://arXiv.org/abs/gr-qc/0412081}. 

 \bibitem{schutz} B.F. Schutz, {\it Geometrical methods of
 mathematical physics}, Cambridge, 1980.

 \bibitem{schroer}B. Schroer,
 {\em String theory and the crisis in particle physics
 (a Samisdat on particle physics)},
\url{http://arXiv.org/physics/0603112}.
Russian translation of this paper (including its
German lyrics) can be found at
\url{http://th1.ihep.su/soloviev/perevod/schroert.pdf}.

\bibitem{auon}
E. Ay\'{o}n-Beato, A. Garc\'{i}a,
\emph{Phys. Lett. B} \textbf{464}, 25 (1999).

\bibitem{nashed}
Gamal G.L. Nashed, {\it Regular Charged Solutions
in Teleparallel Theory of Gravity}, \\
\url{http://arXiv.org/abs/gr-qc/0610058}.

\bibitem{po} L.S. Pontryagin,
{\it Smooth manifolds and their applications in homotopy
theory}, Am.\ Math.\ Soc.\ Trans. Ser.~2,  {\bf 11},  1--114
(1959).

\bibitem{du}   
  B.A. Dubrovin, A.T. Fomenko and S.P. Novikov, {\it Modern Geometry
  -- Methods and Applications},
 Springer-Verlag, 1984.

\bibitem{blak}
A.L. Blakers, W.S. Massey,  {\it The homotopy groups of a triad,
I, II and III.}  Ann.\ Math.\ {\bf 53},  161 (1951); {\bf 55},
   192 (1952); {\bf 58},  409 (1953).

\bibitem{skyrme} T.H.R. Skyrme, Nucl.\ Phys. {\bf 31},
 556 (1962).

\bibitem{rybakov} Yu.\,P. Rybakov,
{\em Stability of Many-Dimensional Solitons in Chiral Models
and Gravitation.} // VINITI series ``Classical Field Theory and
Gravitational Theory''. V.\,2.
Gravitation and Cosmology. Moscow: VINITI,
1991. P.56-111.

\bibitem{ward}
 R.S. Ward, Phys.\ Rev. {\bf D70}, 061701 (2004);
 \url{http://arXiv.org/hep-th/0407245}.

\bibitem{fjo}F.I. Fjodorov, {\em Lorentz group},
Moscow, Science, 1979. In Russian.

\bibitem{the}I.L. Zhogin,
 {\it Research in theory of Riemannian space with Absolute
 Parallelism.}
 Ph.D.\ Thesis, Tomsk University, May 1996
 (Rus; see also \ \url{http://arXiv.org/gr-qc/0412130}).

\bibitem{zh5}I.L. Zhogin,
{\it Singularities in a gravitation theory}.
 J. Russian Phys. {\bf 35}, 647 (1993).

\bibitem{aspect}
A. Aspect, J. Dalibard, and G. Roger, Phys.\ Rev.\ Lett.
{\bf 49}, 1804 (1982).

\bibitem{weihs}
G. Weihs, T. Jennewein, C. Simon, H. Weinfurter, and A.
Zeilinger, Phys. Rev. Lett. {\bf 81}, 5039 (1998);
\url{http://arXiv.org/quant-ph/9810080}.

\bibitem{tit}
W. Tittel, G. Weihs, {\it Photonic Entanglement
for Fundamental Tests and Quantum Communication},
\url{http://arXiv.org/quant-ph/0107156}.

\bibitem{sites} See links: \
\texttt{http://departments.colgate.edu/
physics/research/%
Photon/root/}\\
{\tt photon\_quantum\_mechanics.htm}\,; \
\url{http://marcus.whitman.edu/~beckmk/QM/}\,.

\bibitem{nik}
H. Nikoli\'{c}, {\it Quantum mechanics: Myths and facts}, \
\url{arXiv.org/quant-ph/0609163}.

\bibitem{santos}
E. Santos, {\it
Bell's theorem and the experiments: Increasing empirical
support to local realism?}
\url{arXiv.org/quant-ph/0410193}.

\bibitem{khr}
G. Adenier, A. Khrennikov,
{\it Anomalies in experimental data for the EPR-Bohm experiment:
 Are both classical and quantum mechanics wrong?},
\url{arXiv.org/quant-ph/0607172};
{\it Is the Fair Sampling Assumption supported by EPR
Experiments?}, \\
\url{http://arXiv.org/quant-ph/0606122}.

\bibitem{gis}
N. Gisin, {\it  How come the Correlations?},
\url{http://arXiv.org/quant-ph/0503007}.

\bibitem{fel}
J. L. Fellows,
{\it  Untangling Quantum Entanglement}, \
\url{arXiv.org/quant-ph/0511134}.

\bibitem{rov}
C. Rovelli, {\it Unfinished revolution}, \
\url{http://arXiv.org/gr-qc/0604045}.

\bibitem{marshall}
W. Marshall, C. Simon, R. Penrose, D. Bouwmeester,
{\em Towards quantum superpositions of a mirror},
Phys. Rev. Lett. {\bf 91}, 130401 (2003);
\url{http://arXiv.org/quant-ph/0210001}.

\end{thebibliography}
\end{document}